# Charge Separation in the Photosystem II Reaction Center Resolved by Multispectral Two-Dimensional Electronic Spectroscopy

Hoang H. Nguyen,[1] Yin Song[1,2], Elizabeth L. Maret[1], Yogita Silori[1], Rhiannon Willow[1], Charles F. Yocum[3] and Jennifer P. Ogilvie[1]

[1]Department of Physics and Biophysics, University of Michigan, 450 Church St, Ann Arbor, MI 48109, USA
[2] Beijing Institute of Technology, 5 Zhongguancun South Street, Haidian District, Beijing, China
[3]Department of Molecular, Cellular and Developmental Biology and Department of Chemistry, University of Michigan, 450 Church St, Ann Arbor, MI 48109, USA
*jogilvie@umich.edu*

**Abstract:**

The photosystem II reaction center (PSII-RC) performs the primary energy conversion steps of oxygenic photosynthesis. While the PSII-RC has been studied extensively, the similar timescales of energy transfer and charge separation, and the severely overlapping pigment transitions in the $Q_y$ region have led to multiple models of its charge separation mechanism and excitonic structure. Here we combine two-dimensional electronic spectroscopy (2DES) with a continuum probe and two-dimensional electronic vibrational spectroscopy (2DEV) to study the cyt b559-D1D2 PSII-RC at 77K. This multispectral combination correlates the overlapping $Q_y$ excitons with distinct anion and pigment-specific $Q_x$ and mid-IR transitions to resolve the charge separation mechanism and excitonic structure. Through extensive simultaneous analysis of the multispectral 2D data we find that charge separation proceeds on multiple timescales from a highly delocalized excited state via a single pathway in which $Pheo_{D1}$ is the primary electron acceptor, while $Chl_{D1}$ and $P_{D1}$ act in concert as the primary electron donor.

**Teaser:** Through multispectral 2D electronic spectroscopy we resolve the mechanism of charge separation and the excitonic structure that underlies the primary energy conversion steps of oxygenic photosynthesis.

# Introduction:

The primary charge separation steps of oxygenic photosynthesis occur within Photosystem II (PSII), the only known natural enzyme capable of light-driven water-splitting. The unique capabilities of PSII motivate a deep understanding of its structure-function relationship and highly efficient charge separation mechanism. In plants and green algae, PSII is composed of over 25 subunits; about 250 chlorophyll molecules act as light-harvesting antennas to transfer energy to the reaction center, which performs the initial charge separation. The PSII reaction center (PSII RC), contains six chlorophyll a (Chl *a*) molecules, two pheophytin (Phe), two β-carotenes (Car) and two quinones arranged along the D1 and D2 branches (depicted in Fig. 1A), with charge separation occurring along the D1 side with near-unity quantum efficiency (*1*). The mechanism of charge separation in the PSII RC has been a topic of debate for many years. Initially, a "multimer model" treated all pigment site energies as equal, and suggested that "P680, the primary electron donor of PS II, should not be considered a dimer but a multimer of several weakly coupled pigments, including the pheophytin electron acceptor" (*2*). This model drew an analogy with the better understood purple bacterial reaction center in which the primary charge separation occurs between the strongly-coupled "special pair" pigments (*3*), but concluded that the weaker relative coupling between the $P_{D1}$−$P_{D2}$ pair in the PSII RC would lead to a higher degree of delocalization. Later exciton models assigned distinct site energies to the pigments (*4-10*), and proposed alternative models of charge separation.

Ultrafast spectroscopic measurements have played a key role in motivating and testing exciton and charge separation models of the PSII RC, with early work summarized in review papers (*11-13*). Despite the extensive studies of the PSII RC the similar timescales of energy transfer and charge separation in this system, and the severely overlapping pigment transitions in the $Q_y$ region have led to ongoing debate about its charge separation mechanism and excitonic structure. Two main charge separation mechanisms have been proposed for the PSII RC. Based on a photon echo study, Prokhorenko and Holzwarth proposed that the accessory chlorophyll on the D1 branch ($Chl_{D1}$) is the primary electron donor in the charge separation process (the "$Chl_{D1}$ pathway": RC*→ $Chl_{D1}^+Pheo_{D1}^-$ → $P_{D1}^+Pheo_{D1}^-$ ) (*14*). Alternatively, transient absorption studies by Shuvalov and coworkers claim that charge separation proceeds in a manner similar to the purple bacterial reaction center, in which charge separation begins at the "special pair" and the accessory $Chl_{D1}$ acts as the primary electron acceptor (the "$P_{D1}$ pathway": RC*→ $P_{D2}^+P_{D1}^-$ → $P_{D1}^-Chl_{D1}^+$ → $P_{D1}^+Pheo_{D1}^-$))(*15*). Additional spectroscopic measurements including transient absorption measurements with broadband visible and mid-infrared probes as well as recent two-dimensional electronic spectroscopy experiments have supported either the $Chl_{D1}$ hypothesis (*16, 17*), or have proposed that charge separation proceeds via both the $Chl_{D1}$ and $P_{D1}$ pathways (*18, 19*).

In order to extract the richest information content from the system with simultaneous high frequency and temporal resolution, we utilize Fourier transform two-dimensional (2D) spectroscopy (*20*), employing a combination of probes in the visible and mid-infrared region as depicted in Fig. 1B. Previous 2D electronic spectroscopy (2DES) experiments of the PSII-RC by our group (*21, 22*) and others (*23-27*) have focused on the $Q_y$ region where the severe overlap of the excitons complicates the interpretation of the data and hinders tests of the exciton model and charge separation mechanism. While exciting within the crowded $Q_y$ region is key to unraveling the sequence of events that proceed from the population of specific $Q_y$ excitonic states, it is critical to obtain pigment-specific spectroscopic signatures to distinguish between distinct charge separation pathways. Previous transient absorption spectroscopy employing broadband probes have exploited the improved spectral separation of Chl and Pheo $Q_x$ transitions, and the formation of distinct anion absorption bands for this purpose (*28, 29*). Additionally, the C=O keto and ester vibrational bands of Chl and Pheo have been shown to be sensitive to the protein environment and to charge separation, and have been used by Groot et al. in transient absorption studies of the PSII RC with a mid-IR probe (*16*). The approach of combining 2D electronic excitation with a mid-IR probe (2DEV spectroscopy (*30*)) was recently employed by Yoneda et al. to study the PSII RC (*31*). Here we combine broadband 2DES (*32*) and 2DEV (*33*), exciting the $Q_y$ band to interrogate its excitonic structure and initiate charge separation. We probe over the visible and mid-IR regions to access a unique combination of pigment-specific spectroscopic markers and use lifetime density analysis and global target analysis with simultaneous fitting of the broadband-2DES and 2DEV data to test models of charge separation. We observe and characterize delocalized excitonic and charge transfer (CT) and states, as well as trap states (*34*) that are responsible for slower phases of charge separation. The combination of visible and mid-IR probes reveal a highly delocalized charge separation process in which $Pheo_{D1}$ acts as the primary electron acceptor, with $Chl_{D1}$ and $P_{D1}$ acting in concert as the primary electron donor. This mechanism unites the distinct $Chl_{D1}$ and $P_{D1}$ pathways that have been proposed based on lower-dimensional measurements that probed a subset of the transitions considered here.

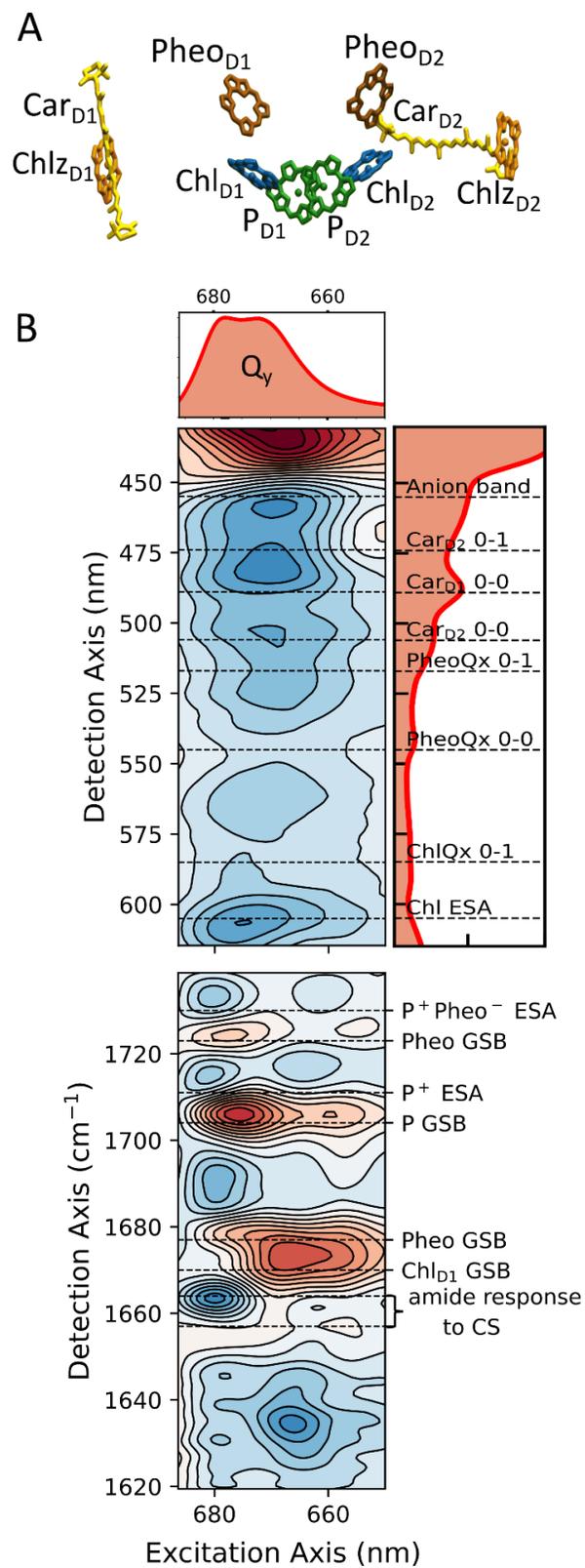

**Figure 1A.** Arrangement of the PSII-RC pigments, based on the 1.9 Å structure (PDB:3WU2) of *T. vulcanus* (*35*), **B.** Representative multispectral 2D spectroscopy used in this study, where we combine broadband 2DES and 2DEV spectra (shown here at $t_2$=200 fs). The 2D excitation axis spans the spectrally congested $Q_y$ region, while broadband detection in the visible (2DES) and mid-IR (2DEV) accesses transitions specific to the RC pigments and charge-separated species.

A 2D spectroscopy experiment utilizes a sequence of three ultrafast laser pulses with carefully controlled inter-pulse timing to induce a third order polarization in the interrogated sample (*20*). The first two laser pulses (the pump pulses) separated by a coherence time $t_1$ generates an excited state or ground state population, which decays over the "waiting time" $t_2$ until the arrival of the third pulse (the probe pulse). The signal radiated as a result of the third order polarization generated in the sample can be detected in the frequency domain with a spectrometer, defining the detection frequency axis. Fourier transformation of the signal with respect to $t_1$ generates the excitation axis. 2D correlation plots of detection versus excitation frequency can be generated at a series of $t_2$ waiting time points to map out the excited-state evolution of the system.

## Results:

Figure 2A shows the waiting time dependence of the broadband-2DES data. Upon $Q_y$ excitation, the dominant features of the 2DES spectra are an initial positive ground state bleach (GSB) signature in the Soret band, and a broadband negative excited state absorption (ESA) signal spanning the $Q_x$ and carotenoid regions (*13, 36*). Charge separation is evident from the increasingly negative signal at 455 nm at longer waiting times, corresponding to the formation of ESA of the Pheo$^-$ anion (*18, 37, 38*). Also evident is the expected increase in Pheo $Q_x$ 0-0 GSB at 545 nm that appears after a few picoseconds and continues to grow more positive at long waiting times, consistent with formation of a charge separated species involving Pheo.

The waiting time dependence of 2DEV data is shown in Figure 2B. Compared to the broadband-2DES spectra, the 2DEV spectra show considerably more pronounced excitation frequency dependence. This is likely due to the lack of broad ESA that dominates the 2DES spectra, obscuring the excitation-dependent features that are more clearly observed upon approximate subtraction of the ESA signal (see Supplementary Figure S1B). The excitation-dependence in both the 2DES and 2DEV spectra can be roughly partitioned into two distinct regions centered at ~665 and ~680 nm with rich and distinct waiting time-dependent structure. We base our assignment of the mid-IR features on the work of Nabedryk et al. (*39*), Noguchi et al. (*40, 41*) and Groot et al. (*16*) as summarized in the Supplementary Information (Section S2). Upon excitation at 680 nm, we observe immediate GSB of Pheo via positive peaks at 1722 cm$^{-1}$ and 1677 cm$^{-1}$ (*39, 41*), as well as GSB of P via a positive peak at ~1704 cm$^{-1}$ (*16, 41*). A broadband negative signal is present across the mid-IR detection window, with the initial strongest contribution at ~1664 cm$^{-1}$, which in combination with the positive feature at 1657 cm$^{-1}$ has been attributed to the amide C=O response to charge separation (*16, 39, 41*). Also notable is a negative feature at ~1711 cm$^{-1}$ that is present at $t_2$=200 fs and grows in amplitude that been attributed to the keto C=O of P$^+$ (*16, 41*). Upon excitation of the higher energy $Q_y$ excitons (centered at 665 nm) the dominant initial features include a broad positive GSB feature spanning ~1670-1680 cm$^{-1}$, encompassing assigned Chl$_{D1}$ and Pheo GSB features at 1670 cm$^{-1}$ and 1677 cm$^{-1}$ respectively (*16, 39*). Also present, although with smaller amplitude than seen with 680 nm excitation, is the P GSB at 1704 cm$^{-1}$. We note that the negative peak we observe near 1635 cm$^{-1}$ at early times is present in Pheo$^-$-minus-Pheo light-induced FTIR difference spectra of dithionite-treated samples (*39*).

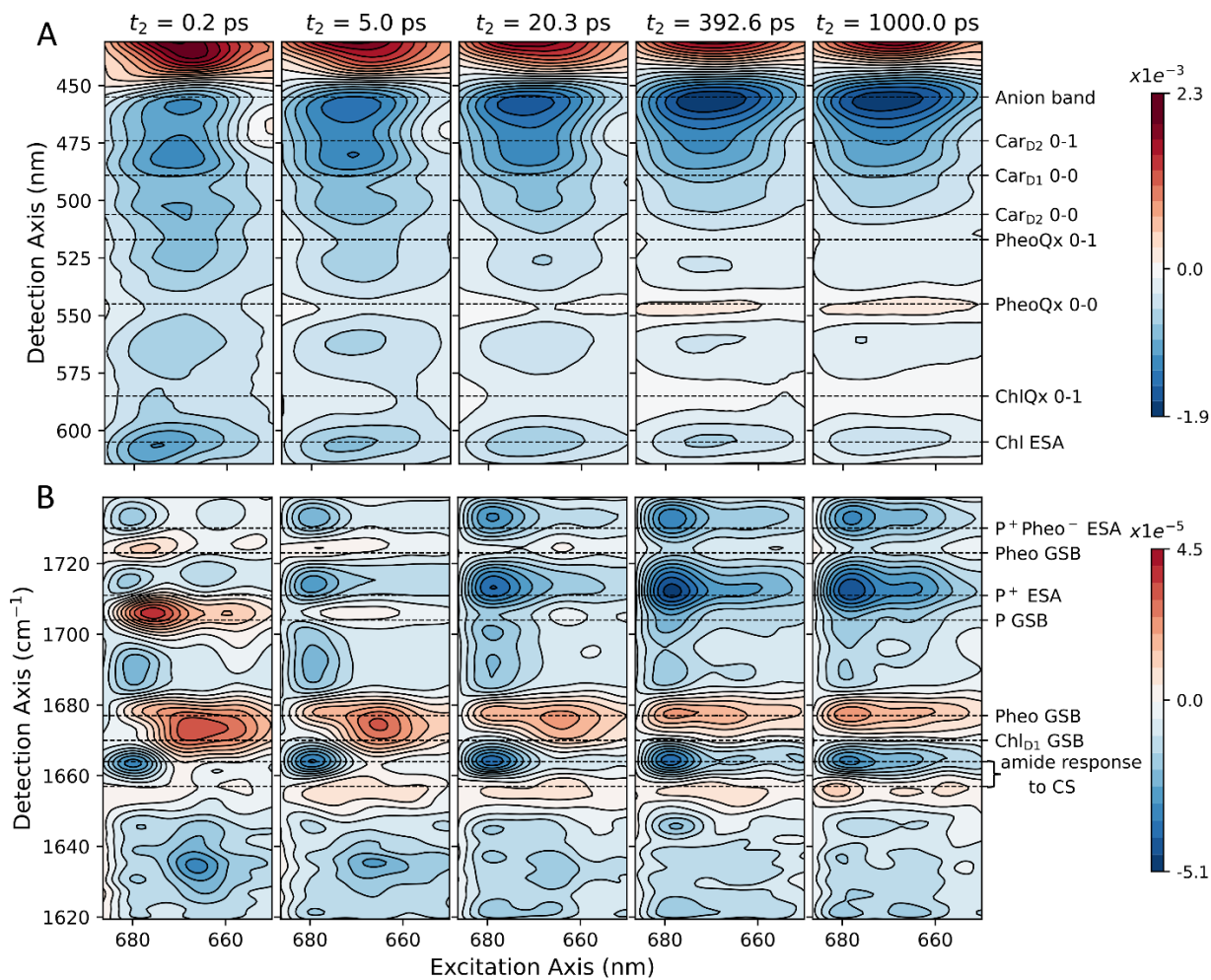

**Figure 2A.** Broadband-2DES spectra plotted at various t$_2$ times. The dashed lines denote specific spectral features in the Q$_x$ region: Chl Q$_x$ 0-1 at 585 nm, Pheo Q$_x$ 0-0 at 545 nm, Pheo Q$_x$ 0-1 at 512 nm, Car$_{D2}$ 0-0 at 506 nm, Car$_{D2}$ 0-1 at 474 nm, and Car$_{D1}$ 0-0 at 489 nm. The spectral locations of these peaks are taken from the linear absorption analysis done by Shipman et al. (*42*) and Rätsep et al. (*43*). **B.** 2DEV spectra at corresponding t$_2$, with assignments derived from the literature (*16, 39-41*). Frequencies of notable transitions are denoted by horizontal dashed lines. The 2DEV spectra shows clearer excitation frequency dependence.

**Data Analysis**

To obtain an overview of the kinetics in the broadband 2DES and 2DEV data sets we applied lifetime density analysis (LDA) (*29, 44*). LDA fits the data using exponential functions with a continuous distribution of time constants. As no prior knowledge of the system dynamics is required for this analysis, this approach is well-suited for studying complicated systems such as the PSII-RC. We further process the lifetime density maps (LDMs) obtained from the LDA (implemented via the OPTIMUS software package (*44*)) using a convolution method to reduce unphysical oscillatory components following reference (*45*) and personal communication with Mark Berg at U of SC. Details of this treatment are given in the Supplementary Information (Section S3).

Figure 3 displays the lifetime density maps (LDMs) obtained from integration of the 2D spectra within two different excitation windows centered at 680 (top row) and 665 nm (bottom row) to capture the main excitation-frequency-dependent kinetics present in the broadband 2DES (left column) and 2DEV (right column) data. In the LDMs, blue features indicate a rise of GSB or decay of ESA, while conversely, red indicates decay of GSB or rise of ESA. Direct comparison of the kinetics revealed by the LDMs of the broadband 2DES and 2DEV data show similar overall waiting-time-dependent evolution of the spectra, consistent with the expectation that the electronic and vibrational probes provide complementary views of the energy conversion process in the PSII RC. Within both excitation windows, spectral changes can be roughly characterized as taking place with distinct processes within the first ~1ps, ~1-100 ps and >100ps. Additional LDMs derived from the 2D data in 3 nm excitation windows spanning the $Q_y$ region are provided in the Supplementary Information.

*"Red" excitation (677-683 nm):* Upon red excitation, the broadband 2DES reveals immediate signatures of charge separation as indicated by the 455 nm anion ESA and the broadband ESA features throughout the visible and mid-IR. The initial rise in broadband ESA, most apparent in the 2DEV LDM, is complete within ~0.3 ps. At ~0.3 ps, there is an overall decrease of the GSB of P at 1705 cm$^{-1}$ which at later times is difficult to distinguish from the rising ESA from P$^+$ at 1711 cm$^{-1}$. At ~3 ps the 2DES LDM exhibits a peak in the anion ESA coincident with a rise in Pheo GSB, consistent with Pheo signatures in the 2DEV LDM on this timescale. Both 2DES and 2DEV LDMs exhibit a later stage of charge separation on the timescale of >100 ps as evident from the strong growth in the anion ESA at 455nm and the P$^+$ ESA at 1711 cm$^{-1}$, accompanied by Pheo GSB growth and positive/negative features at 1657/1664 cm$^{-1}$ associated with the amide C=O response to charge separation (*16, 39, 41*).

*"Blue" excitation (662-668 nm):* Upon blue excitation, the 2DES LDM reveals that the initial charge separation is much weaker than upon red excitation, as indicated by a relatively small initial anion band absorption and Pheo GSB. The 2DEV LDM also shows early Pheo GSB and a decay of P GSB. Between ~3-20 ps the growth of Pheo GSB competes with ESA from P$^+$Pheo$^-$ and P$^+$ indicating increasing charge separation on this timescale. We note the concomitant bandshift feature present in the Car region on this timescale, consistent with previous reports from transient absorption studies (*46*). The strong broad feature between 1660-1680 cm$^{-1}$ at ~20 ps has contributions from the C=O response to charge separation (*16, 39, 41*). The GSB features in this region are difficult to interpret due to their close proximity and competing oppositely-signed processes occurring on this timescale. On the timescale of >100 ps, both 2DES and 2DEV LDMs exhibit a later stage of charge separation, with spectral markers that are consistent with the red excitation data.

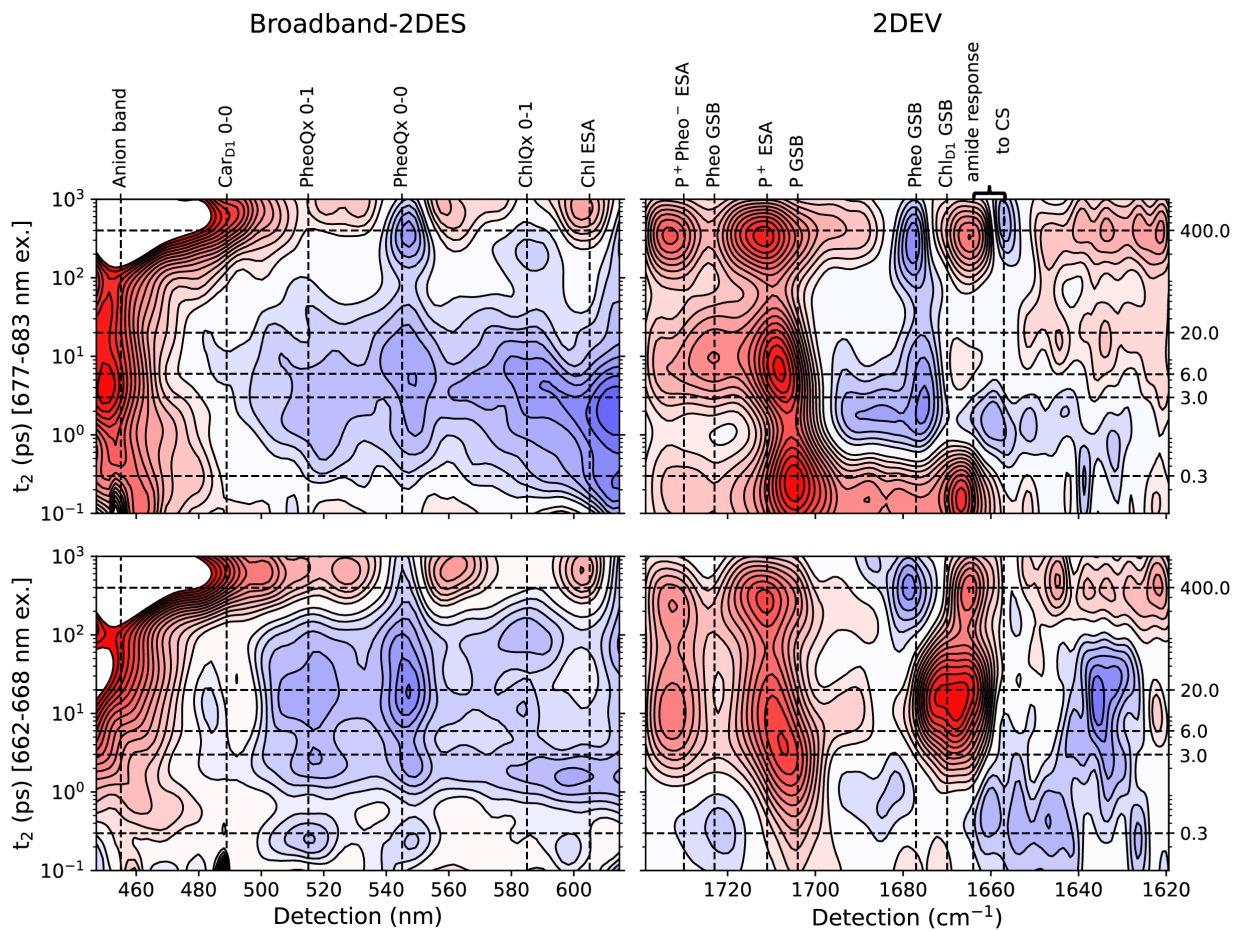

**Figure 3** LDMs of broadband-2DES (left) and 2DEV (right) at different excitation frequency regions: 677-683 nm (top) and 662-668 nm (bottom). Blue features indicate a rise of GSB or decay of ESA, while conversely, red indicates decay of GSB or rise of ESA.

*Global Target Analysis:* To further explore the charge separation mechanism and establish a model that can capture the kinetic processes in the multispectral data we turn to global-target analysis (*47*). This approach fits the data using a trial kinetic model and produces a set of species associated spectra (SAS) and rate constants. Here we perform simultaneous fits of broadband-2DES and 2DEV data to multiple kinetic models, chosen based on the timescales and spectral features obtained in the LDMs as well as models from the literature. We extensively tested the two-pathway model of charge separation as well as other models as discussed in the Supplementary Information. We found that the two-pathway model proposed by Romero et al.(*18*) did not produce distinct SAS consistent with a distinct $P_{D1}$ pathway that lacks participation from Pheo. While a simplification of the two-pathway model that omitted the slow energy transfer from Chlz pigments was found to adequately fit the broadband 2DES data, it failed to give reasonable results for the 2DEV fits. The best model that adequately describes the kinetics in both broadband 2DES and 2DEV datasets is given in Figure 4A. It consists of five compartments: two distinct exciton states (RC* and Trap*), two intermediate Radical Pair (RP1 and RP2), and the final charge separated state $P_{D1}^+Pheo_{D1}^-$ (RP3). The lifetime of the final charge separated state RP3 is much longer than the ~1 ns waiting time limit of our experiment. We therefore fix the RP3 lifetime to 13 ns following the work of Romero et al. (*28*). Within our model a rapid charge separation channel exists: RC* → RP1 → RP2, followed by a slow ~150 ps relaxation process of the charge separated state RP2 → RP3. A slower charge separation route also proceeds from a distinct trap state. The existence of trap states at 77 K (*18*), 4 K (*34, 48*) and room temperature (*17*) have been previously proposed, as has a slow phase associated with the formation of the relaxed charge separated state (*18*). We find that Trap* is depopulated within ~13 ps, proceeding to form RP2 via the same pathway as RC*. The 2D-SAS of the visible and IR show consistent excitation wavelength dependence, lending credibility to our target model through which the final charge separated state is reached upon excitation of two distinct exciton states (RC* and Trap*).

The 2D SAS in the mid-IR exhibit some spurious structure in regions of low amplitude, motivating a simpler representation that captures the excitation-frequency-dependence and enables visualization of the spectroscopic features associated with each compartment of our target model. For this purpose, we make the approximation that the 2D SAS of the two initial excitons (RC*, Trap*) can be represented as a product of a 1D SAS and a Gaussian distribution that captures the excitation frequency dependence of the 2D data. The resulting Gaussian distributions and 1D SAS are shown in Figures 5 and 6 respectively. The fitting procedure is described in further detail in the Supplementary Information (Section S6), where we show that this reduced representation reproduces the main 2D spectral features and the kinetic processes present in both data sets.

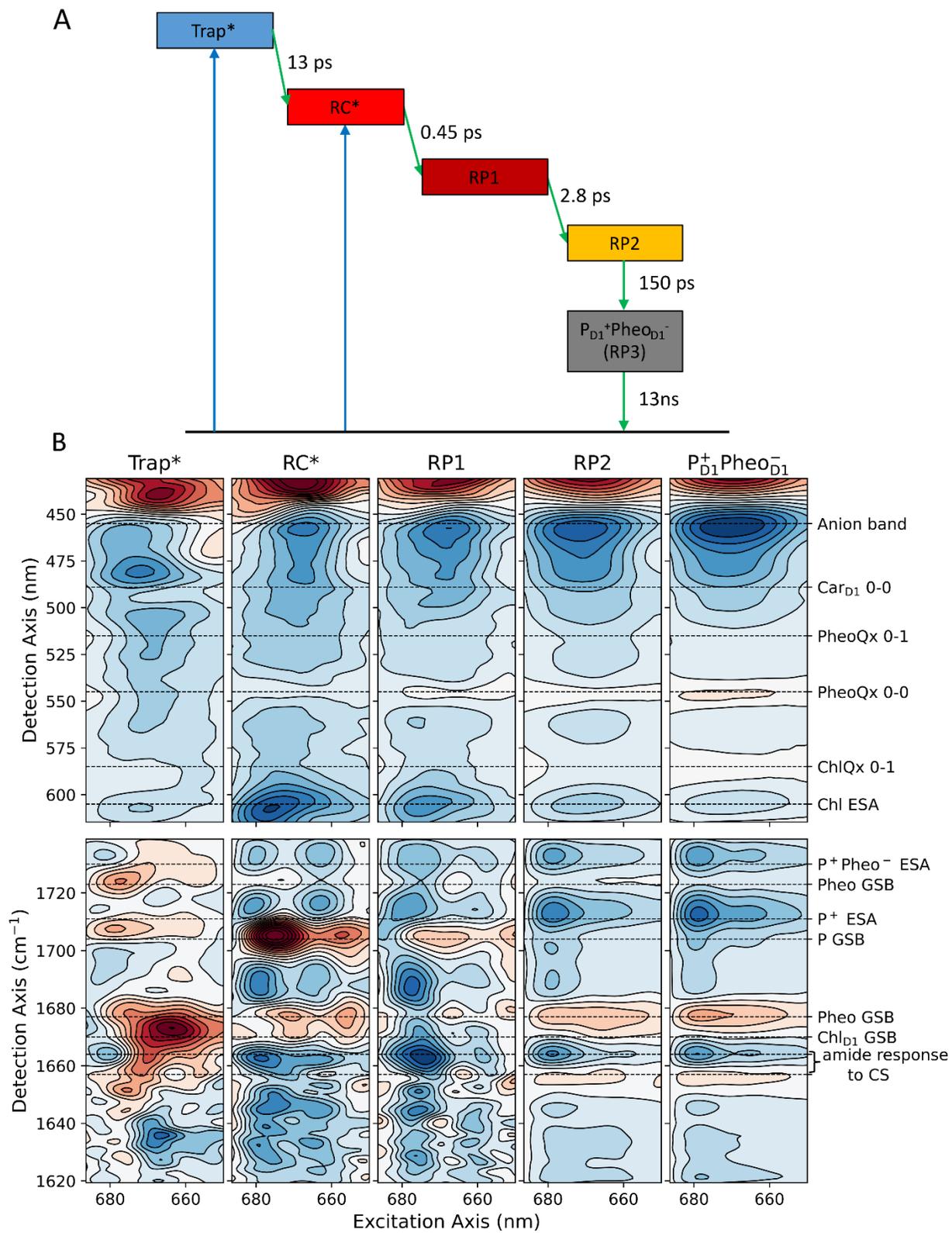

**Figure 4** Result from simultaneously fitting broadband-2DES and 2DEV data: **A.** 5-compartment model with 1 trap state and 3 RP states, **B.** 2D SAS of each compartment, with both visible (top) and mid-IR (bottom) components.

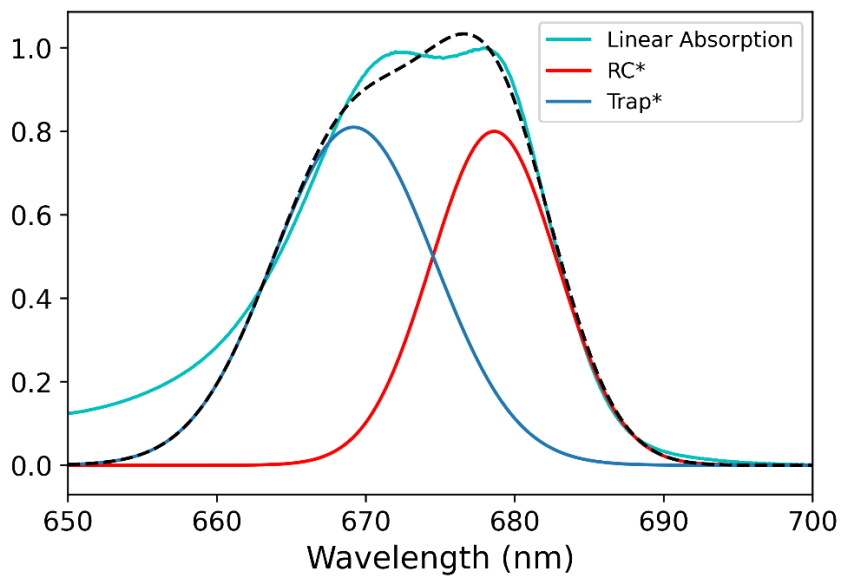

**Figure 5** Linear absorption of the D1D2 RC at 77 k (green solid curve) fit with three Gaussian distributions that represent the two distinct initial exciton states RC* (red solid curve, centered at 679 nm, full width at half maximum (FWHM) 4.2 nm), and Trap* (blue solid curve, centered at 669 nm, FWHM 5.4 nm). The black dashed line is the sum of the two Gaussian distributions.

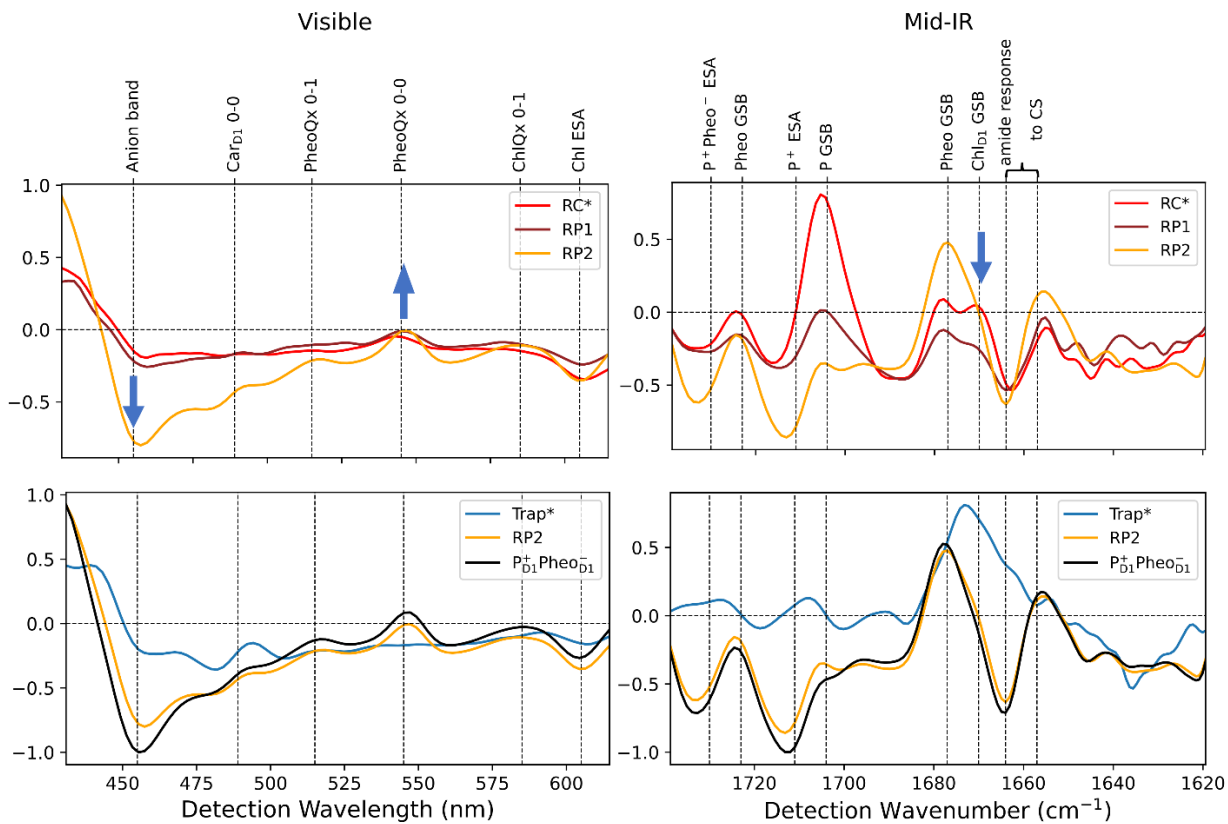

**Figure 6** 1D SAS of the compartments in the visible (right) and mid-IR (left). The amplitudes are normalized with respect to the final charge separated state RP3 ($P_{D1}^+Pheo_{D1}^-$).

## Discussion:

The 1D SAS shown in Figure 6 allow us to clearly visualize the spectroscopic features associated with each compartment of the global target model. First, we examine the rapid charge separation process that proceeds upon excitation of RC*, with excitation centered at 679 nm. We then discuss the slower charge separation proceeding from the trap states and the implications for our understanding of the mechanism of charge separation in the PSII RC.

*Rapid Charge Separation:* The visible SAS for RC* exhibits GSB of the Pheo $Q_x$ peaks as well as ESA from Chl ESA and the Pheo anion band. In the mid-IR SAS, we observe Pheo, $Chl_{D1}$ and P GSB peaks. Thus RC* is consistent with a delocalized exciton state spanning the RC pigments: $(Chl_{D1}Pheo_{D1}P)^*$. This state also appears to have some degree of CT character, consistent with Stark spectroscopy experiments that propose direct excitation of mixed exciton-CT states $(Chl_{D1}^{\delta+}Phe_{D1}^{\delta-})^*$ at 681nm, and $(P_{D2}^+P_{D1}^-)^{\delta *}$ at 684nm (*49*). In the visible SAS, from RC* to RP1, we see a rise of Pheo $Q_x$ GSB and Pheo anion ESA, signatures of charge separation involving Pheo$^-$. The Pheo $Q_x$ peaks also experience a slight Stark shift. In the mid-IR SAS, a GSB peak of $Chl_{D1}$ at 1670 cm$^{-1}$ is present in the 1D SAS of RC* and RP1, but not RP2 or RP3. The GSB peak associated with P decreases significantly from RC* to RP1, accompanied by an increase in ESA from $P^+$. These observations are consistent with assignment of RP1 as $(Chl_{D1}P_{D1})^+Pheo_{D1}^-$. Comparing RP2 to RP1, the visible SAS reveal an increase in the Pheo anion band and a slight increase in the Pheo GSB with an additional small Stark shift. The Chl ESA is also increased in comparison with RP1. In the mid-IR a comparison of the SAS of RP1 and RP2 shows an increase in Pheo GSB, as well as $P^+$ and $Pheo^-$ ESA. The reduced P GSB peak in the mid-IR is likely a result of the competing nearby $P^+$ ESA. RP2 also exhibits the expected strong derivative-like amide response that has been associated with charge separation. The visible and mid-IR SAS support the assignment of RP2 as the final charge separated state $P_{D1}^+Pheo_{D1}^-$. We note a slight red shift in the $P^+$ ESA peak from RC* to RP1 to RP2, which Yoneda et al. assigned to the hole migration (*31*), consistent with our assignments. The mid-IR SAS for RP2 and RP3 are very similar and are in good agreement with the final radical pair spectrum reported by Groot et al. in their pump mid-IR probe study of the PSII RC (*16*). Based on the combination of visible and mid-IR SAS we find that a consistent picture in which the charge separation in the PSII RC proceeds via the following sequential pathway:

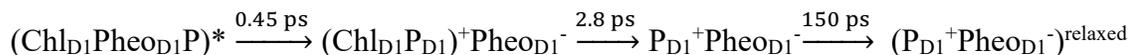

$$(Chl_{D1}Pheo_{D1}P)^* \xrightarrow{0.45\ ps} (Chl_{D1}P_{D1})^+Pheo_{D1}^- \xrightarrow{2.8\ ps} P_{D1}^+Pheo_{D1}^- \xrightarrow{150\ ps} (P_{D1}^+Pheo_{D1}^-)^{relaxed}$$

The timescales are in reasonable agreement with previous works which suggest that initial charge separation at 77 K occurs within 600–800 fs (*16*) or ~1–3 ps (*18*), followed by a slower relaxation process that has also been reported in both room temperature and 77 K experiments (*17, 18*).

*Charge Separation from the Trap state:* Consistent with previous reports (*17, 18, 37*), we also observed slower phases of charge separation that we interpret as originating from trap states. Trap*, excited at ~669 nm, exhibits visible SAS features that include Chl ESA, Chl $Q_x$ bleach, and a derivative-like feature in the Car region. Notably no Pheo $Q_x$ bleach or Pheo anion band signatures are present. The mid-IR features of Trap* are highly distinct from the other SAS, with a broad prominent GSB centered at ~1673 $cm^{-1}$ and an ESA at ~1635 $cm^{-1}$. The broad GSB encompasses several identified mid-IR bands, complicating its assignment. Based on the visible SAS and the lifetime of ~20 ps (*28, 29*), we propose that Trap* represents the Chlz pigments which have been reported to transfer energy to the center pigments on this timescale, accompanied by a Stark shift of the Car peaks (*50*). We note that FTIR studies of spinach membranes report 80 K absorption from neutral Chlz at 1676 $cm^{-1}$ and 1687 $cm^{-1}$ (*51*) which is reasonably consistent with our Trap* SAS. We note that a modified global target model that produces a comparable fit to the data and very similar SAS has Trap* decay directly to RP2 on a similar timescale (~24 ps) as described in the Supplementary Information. We also considered an additional model with two trap states, similar to that proposed by Romero et al.(*18*) to account for the slowest charge separation processes evident in the LDMs. This model is presented in the Supplementary Information and includes a Trap2* exciton at ~670nm. While this model provided a reasonable fit to the data, it appeared unphysical: the visible and mid-IR SAS were inconsistent, and the mid-IR SAS closely resembled RP3, suggesting that RP3 is photoexcited directly, with little subsequent evolution.

Theoretical studies of the PSII RC are highly challenging due to the number of pigments and the necessity of accounting for the effect of the protein environment. Recent high-level QM/MM calculations find that $Chl_{D1}$ has the lowest site energy and support its assignment as the primary electron donor (*7, 52*). The calculation of CT states is particularly difficult, and it is our hope that our spectroscopic measurements will provide feedback for testing and refining theoretical models of the PSII RC. Our findings are in reasonable agreement with the empirical exciton models of Novoderezhkin (*9*), Gelzinis (*10*) and Renger (*6*). The Novoderezhkin model has high participation from $Chl_{D1}$, $Pheo_{D1}$ and $P_{D1}$ pigments in the lowest energy $Q_y$ exciton states, consistent with our RC* exciton assignment of $(Chl_{D1}Pheo_{D1}P)*$ that is excited on the red edge (~679 nm). In both the Gelzinis and Renger models the lowest energy exciton state is dominated by $Chl_{D1}$. Our identification of Trap* as the Chlz pigments is consistent with the Novoderezhkin and Gelzinis models, in which they participate heavily in the highest energy $Q_y$ excitons (*9, 10*).

To date, there has been little consensus on the charge separation mechanism of the PSII RC. Transient absorption measurements exciting the $Q_y$ region and probing in the visible or mid-IR have favored either the $Chl_{D1}$ (*16, 17*) or $P_{D1}$ pathways (*15*). Still others have suggested that both pathways occur, with protein conformation dictating which pathway will be taken (*18*). A 2DES study in the $Q_y$ region used global analysis to find support for the $Chl_{D1}$ pathway (*19*). Based on calculations they proposed that, while the $Chl_{D1}$ pathway was dominant, the $P_{D1}$ pathway was viable but not well resolved in the experiments due to its minor relative contribution. Here we have used 2D excitation spanning the $Q_y$ region combined with a multispectral probe to access a broad range of pigment-specific and anion/ion spectral markers spanning the visible and mid-IR. Without relying on a global target model, the LDMs demonstrate that sub-picosecond spectral signatures associated with $Pheo^-$ are present independent of $Q_y$ excitation frequency (see the Supplementary Information for additional LDMs beyond those shown in Figure 3). This rules out the possibility that a distinct $P_{D1}$ pathway can be accessed via selective excitation of specific $Q_y$ excitonic states. While such a pathway might be accessed on slower timescales following energy transfer from Chlz, our global target analysis failed to produce SAS with consistent visible and mid-IR signatures to support the existence of such a pathway. We note that the spectral resolution of our study with respect to the $Q_y$ excitation frequency axis compares favorably with previous work (*16, 18, 31*), ruling out the possibility that we had insufficient resolution to isolate processes that have been observed in other measurements. Further comparison of our work with other measurements is detailed in the Supplementary Information.

In conclusion, we have used multispectral 2DES, exciting the spectrally congested $Q_y$ region and combining the rich spectroscopic signatures in the visible and mid-IR to gain insight into the $Q_y$ excitonic structure and reveal the charge separation mechanism of the PSII RC. Our measurements resolve excitonic states with CT character within the $Q_y$ region, in reasonable agreement with Stark spectroscopy measurements(*49*). Our extensive kinetic analysis, combining lifetime density and global target analysis of the multispectral data reveals that charge separation in the PSII RC proceeds via a sequential mechanism that borrows elements of the distinct $Chl_{D1}$ and $P_{D1}$ pathways. We find that excitation of a highly delocalized RC exciton $(Chl_{D1}Pheo_{D1}P)*$ leads to rapid charge separation, consistent with the original multimer model which proposed that the precursor to the primary charge separated state in the PSII RC was highly delocalized (*2*). This is analogous to our recent study of the heliobacterial RC (*53*), which has been proposed to be the RC most similar to the common ancestor of all photosynthetic RCs (*54, 55*), in which we observed charge separation from a highly delocalized state via a similar mechanism to that proposed here. Like the $Chl_{D1}$ pathway, and consistent with many previous studies, we find that $Pheo_{D1}$ acts as the primary electron acceptor. However, our combination of visible and mid-IR spectral signatures indicate that the primary electron donor is not strictly $Chl_{D1}$ but involves $Chl_{D1}$ and $P_{D1}$ acting in concert. Such delocalization of the hole in the primary charge separated state would draw cation density away from $Pheo^-$, helping prevent wasteful charge recombination, as we have proposed for the heliobacterial RC (*53*).

## Materials and Methods:

The D1-D2-Cytb559 reaction center (PSII-RC) complexes were extracted from spinach following a method based on the work of Berthold et al. (*56*) and van Leeuwen et al. (*57*). We verified purity of the sample and removal of CP47 and CP43 via a ratio of 1.2 for the linear absorption at 417:435 nm as discussed by Eijckelhoff et al. (*58*). The RC samples were further diluted with glycerol with a ratio of 1:1 (volume : volume). In the broadband-2DES experiment, the RC sample was concentrated to obtain an optical density (OD) of ~1.5 measured at 675 nm at room temperature for a pathlength of 380 μm. In the 2DEV experiment, the OD of the sample measured at 675 nm was ~0.6 at room temperature for a pathlength of 100 μm. Both 2DES and 2DEV measurements were collected at 77 K using an Oxford Instruments cryostat (MicrostatN).

The broadband-2DES and 2DEV experiments were performed at the Laboratory for Ultrafast Multidimensional Optical Spectroscopy (LUMOS) (*59*) at the University of Michigan, utilizing pulse-shaping based 2D spectroscopy in the pump-probe geometry (*60*). The laser repetition rate was set to 500 Hz to avoid a build up of long-lived triplet states in the PSII-RC (*61*). The pump was generated by using a home-built NOPA (*62*) with 50 nJ, and 100 nJ of pulse energy at sample position in the broadband-2DES and 2DEV experiments respectively. The pump was compressed to 12 fs of pulse duration using the SPEAR method (*63*). The $1/e^2$ beam waist of the pump at sample position was ~120 μm, giving a beaching ratio of ~5-6% in both experiments. In broadband-2DES, the probe was a white-light continuum generated by focusing 800 nm light from a regenerative amplifier into a $CaF_2$ crystal. The chirp of the white-light was corrected post data-collection (*64*). In the 2DEV experiment, the mid-IR probe, generated by DFG, was centered at ~5988 nm with a FWHM >300 nm. The mid-IR probe pulse energy was kept below 100 nJ before the sample to avoid both nonlinear processes induced by the probe and saturation of the detector. A two-step phase cycling scheme was used to remove scattering and background signals (*65*). In both experiments, $t_1$ was scanned to a maximum delay of 210 fs. Waiting time $t_2$ was scanned from -1ps to 5 ps in linear steps, then from 5 ps to 1 ns in logarithmically-spaced steps. 21,000 and 693,000 laser shots were used to construct each averaged 2D spectrum in broadband-2DES and 2DEV experiments, respectively. Experiments were repeated at least 3 times on different samples to confirm reproducibility.

## Acknowledgments:

We thank Mark Berg for helpful discussions about lifetime density analysis and for sharing his convolution approach.

**Funding:**

We acknowledge the Office of Basic Energy Sciences, the US Department of Energy grant DE-SC0016384 (HHN, JPO, ELM, RW, YS)

**Author contributions:**

    Conceptualization: JPO

    Methodology: JPO

    Investigation: HHN, YS, ELM, YS, RW

    Sample preparation: HHN, ELM, CFY

    Resources: HHN, JPO

    Writing—original draft: HHN, JPO

**Competing interests:**

The authors declare no competing interests.

**Data and materials availability:**

All data are available in the main text or the supplementary materials.